\documentclass[11pt,a4paper]{article}
\pdfoutput=1
\usepackage{float}
\makeatletter
\gdef\@fpheader{Prepared for submission to JHEP}
\gdef\@journal{jhep}

\newif\ifnotoc\notocfalse
\newif\ifemailadd\emailaddfalse
\newif\iftoccontinuous\toccontinuousfalse

\RequirePackage{amsmath}
\RequirePackage{amssymb}
\RequirePackage{epsfig}
\RequirePackage{graphicx}
\RequirePackage[numbers,sort&compress]{natbib}
\RequirePackage[colorlinks=true
  ,urlcolor=blue
  ,anchorcolor=blue
  ,citecolor=blue
  ,filecolor=blue
  ,linkcolor=blue
  ,menucolor=blue
  ,linktocpage=true
  ,pdfproducer=medialab
  ,pdfa=true
]{hyperref}

\def\@subheader{\@empty}
\def\@keywords{\@empty}
\def\@abstract{\@empty}
\def\@xtum{\@empty}
\def\@dedicated{\@empty}
\def\@arxivnumber{\@empty}
\def\@collaboration{\@empty}
\def\@collaborationImg{\@empty}
\def\@proceeding{\@empty}
\def\@preprint{\@empty}

\newcommand{\subheader}[1]{\gdef\@subheader{#1}}
\newcommand{\keywords}[1]{\if!\@keywords!\gdef\@keywords{#1}\else%
\PackageWarningNoLine{\jname}{Keywords already defined.\MessageBreak Ignoring last definition.}\fi}
\renewcommand{\abstract}[1]{\gdef\@abstract{#1}}
\newcommand{\dedicated}[1]{\gdef\@dedicated{#1}}
\newcommand{\arxivnumber}[1]{\gdef\@arxivnumber{#1}}
\newcommand{\proceeding}[1]{\gdef\@proceeding{#1}}
\newcommand{\xtumfont}[1]{\textsc{#1}}
\newcommand{\correctionref}[3]{\gdef\@xtum{\xtumfont{#1} \href{#2}{#3}}}
\newcommand\jname{JHEP}
\newcommand\acknowledgments{\section*{Acknowledgments}}

\newcommand\preprint[1]{\gdef\@preprint{\hfill #1}}

\newcommand\note[2][]{%
\if!#1!%
\stepcounter{footnote}\footnotetext{#2}%
\else%
{\renewcommand\thefootnote{#1}%
\footnotetext{#2}}%
\fi}

\newcommand\compress{%
\renewcommand\afterProceedingsSpace{\bigskip}%
\renewcommand\afterTitleSpace{\bigskip}%
\renewcommand\afterRuleSpace{\bigskip\bigskip}
\renewcommand\afterEmailSpace{\par\bigskip}}

\newtoks\auth@toks
\renewcommand{\author}[2][]{%
  \if!#1!%
    \auth@toks=\expandafter{\the\auth@toks#2\ }%
  \else
    \auth@toks=\expandafter{\the\auth@toks#2$^{#1}$\ }%
  \fi
}

\newtoks\affil@toks\newif\ifaffil\affilfalse
\newcommand{\affiliation}[2][]{%
\affiltrue
  \if!#1!%
    \affil@toks=\expandafter{\the\affil@toks{\item[]#2}}%
  \else
    \affil@toks=\expandafter{\the\affil@toks{\item[$^{#1}$]#2}}%
  \fi
}

\newtoks\email@toks\newcounter{email@counter}%
\newcommand{\emailAdd}[1]{%
\emailaddtrue%
\ifnum\theemail@counter>0\email@toks=\expandafter{\the\email@toks, \@email{#1}}%
\else\email@toks=\expandafter{\the\email@toks\@email{#1}}%
\fi\stepcounter{email@counter}}
\newcommand{\@email}[1]{\href{mailto:#1}{\tt #1}}

\newcommand*\collaboration[1]{\gdef\@collaboration{#1}}
\newcommand*\collaborationImg[2][]{\gdef\@collaborationImg{#2}}

\newcommand\afterLogoSpace{\smallskip}
\newcommand\afterSubheaderSpace{\vskip3pt plus 2pt minus 1pt}
\newcommand\afterProceedingsSpace{\vskip21pt plus0.4fil minus15pt}
\newcommand\afterTitleSpace{\vskip23pt plus0.06fil minus13pt}
\newcommand\afterRuleSpace{\vskip23pt plus0.06fil minus13pt}
\newcommand\afterCollaborationSpace{\vskip3pt plus 2pt minus 1pt}
\newcommand\afterCollaborationImgSpace{\vskip3pt plus 2pt minus 1pt}
\newcommand\afterAuthorSpace{\vskip5pt plus4pt minus4pt}
\newcommand\afterAffiliationSpace{\vskip3pt plus3pt}
\newcommand\afterEmailSpace{\vskip16pt plus9pt minus10pt\filbreak}
\newcommand\afterXtumSpace{\par\bigskip}
\newcommand\afterAbstractSpace{\vskip16pt plus9pt minus13pt}
\newcommand\afterKeywordsSpace{\vskip16pt plus9pt minus13pt}
\newcommand\afterArxivSpace{\vskip3pt plus0.01fil minus10pt}
\newcommand\afterDedicatedSpace{\vskip0pt plus0.01fil}
\newcommand\afterTocSpace{\bigskip\medskip}
\newcommand\afterTocRuleSpace{\bigskip\bigskip}
\newlength{\affiliationsSep}
\newcommand\beforetochook{\pagestyle{myplain}\pagenumbering{roman}}

\DeclareFixedFont\trfont{OT1}{phv}{b}{sc}{11}

\renewcommand\maketitle{
\pagestyle{empty}
\thispagestyle{titlepage}
\setcounter{page}{0}
\noindent{\small\scshape\@fpheader}\@preprint\par
\afterLogoSpace
\if!\@subheader!\else\noindent{\trfont{\@subheader}}\fi
\afterSubheaderSpace
\if!\@proceeding!\else\noindent{\sc\@proceeding}\fi
\afterProceedingsSpace
{\LARGE\flushleft\sffamily\bfseries\@title\par}
\afterTitleSpace
\hrule height 1.5\p@%
\afterRuleSpace
\if!\@collaboration!\else
{\Large\bfseries\sffamily\raggedright\@collaboration}\par
\afterCollaborationSpace
\fi
\if!\@collaborationImg!\else
{\normalsize\bfseries\sffamily\raggedright\@collaborationImg}\par
\afterCollaborationImgSpace
\fi
{\bfseries\raggedright\sffamily\the\auth@toks\par}
\afterAuthorSpace
\ifaffil\begin{list}{}{%
\setlength{\leftmargin}{0.28cm}%
\setlength{\labelsep}{0pt}%
\setlength{\itemsep}{\affiliationsSep}%
\setlength{\topsep}{-\parskip}}
\itshape\small%
\the\affil@toks
\end{list}\fi
\afterAffiliationSpace
\ifemailadd 
\noindent\hspace{0.28cm}\begin{minipage}[l]{.9\textwidth}
\begin{flushleft}
\textit{E-mail:} \the\email@toks
\end{flushleft}
\end{minipage}
\else 
\PackageWarningNoLine{\jname}{E-mails are missing.\MessageBreak Plese use \protect\emailAdd\space macro to provide e-mails.}
\fi
\afterEmailSpace
\if!\@xtum!\else\noindent{\@xtum}\afterXtumSpace\fi
\if!\@abstract!\else\noindent{\renewcommand\baselinestretch{.9}\textsc{Abstract:}}\ \@abstract\afterAbstractSpace\fi
\if!\@keywords!\else\noindent{\textsc{Keywords:}} \@keywords\afterKeywordsSpace\fi
\if!\@arxivnumber!\else\noindent{\textsc{ArXiv ePrint:}} \href{https://arxiv.org/abs/\@arxivnumber}{\@arxivnumber}\afterArxivSpace\fi
\if!\@dedicated!\else\vbox{\small\it\raggedleft\@dedicated}\afterDedicatedSpace\fi
\ifnotoc\else
\iftoccontinuous\else\newpage\fi
\beforetochook\hrule
\tableofcontents
\afterTocSpace
\hrule
\afterTocRuleSpace
\fi
\setcounter{footnote}{0}
\pagestyle{myplain}\pagenumbering{arabic}
} 

\renewcommand{\baselinestretch}{1.1}\normalsize
\divide\@tempdima\baselineskip
\@tempcnta=\@tempdima
\skip\@mpfootins = \skip\footins
\renewcommand{\@dotsep}{10000}

\newcommand\ps@myplain{
\pagenumbering{arabic}
\renewcommand\@oddfoot{\hfill-- \thepage\ --\hfill}
\renewcommand\@oddhead{}}
\let\ps@plain=\ps@myplain

\newcommand\ps@titlepage{\renewcommand\@oddfoot{}\renewcommand\@oddhead{}}

\numberwithin{equation}{section}

\renewcommand\section{\@startsection{section}{1}{\z@}%
                                   {-3.5ex \@plus -1.3ex \@minus -.7ex}%
                                   {2.3ex \@plus.4ex \@minus .4ex}%
                                   {\normalfont\large\bfseries}}
\renewcommand\subsection{\@startsection{subsection}{2}{\z@}%
                                   {-2.3ex\@plus -1ex \@minus -.5ex}%
                                   {1.2ex \@plus .3ex \@minus .3ex}%
                                   {\normalfont\normalsize\bfseries}}
\renewcommand\subsubsection{\@startsection{subsubsection}{3}{\z@}%
                                   {-2.3ex\@plus -1ex \@minus -.5ex}%
                                   {1ex \@plus .2ex \@minus .2ex}%
                                   {\normalfont\normalsize\bfseries}}
\renewcommand\paragraph{\@startsection{paragraph}{4}{\z@}%
                                   {1.75ex \@plus1ex \@minus.2ex}%
                                   {-1em}%
                                   {\normalfont\normalsize\bfseries}}
\renewcommand\subparagraph{\@startsection{subparagraph}{5}{\z@}%
                                   {1.75ex \@plus1ex \@minus .2ex}%
                                   {-1em}%
                                   {\normalfont\normalsize\itshape}}

\def\fnum@figure{\textbf{\figurename\nobreakspace\thefigure}}
\def\fnum@table{\textbf{\tablename\nobreakspace\thetable}}

\long\def\@makecaption#1#2{%
  \vskip\abovecaptionskip
  \sbox\@tempboxa{\small #1. #2}%
  \ifdim \wd\@tempboxa >\hsize
    \small #1. #2\par
  \else
    \global \@minipagefalse
    \hb@xt@\hsize{\hfil\box\@tempboxa\hfil}%
  \fi
  \vskip\belowcaptionskip}

\renewenvironment{thebibliography}[1]{%
\begin{oldthebibliography}{#1}%
\small%
\raggedright%
\setlength{\itemsep}{5pt plus 0.2ex minus 0.05ex}%
}%
{%
\end{oldthebibliography}%
}
\makeatother
\usepackage{booktabs,array,mathtools,bm,slashed,microtype}
\usepackage{tikz,pgfplots}
\usepackage{xcolor,longtable}
\usetikzlibrary{arrows.meta,decorations.markings,decorations.pathmorphing,calc,shapes.geometric}
\usepgfplotslibrary{groupplots}
\pgfplotsset{compat=1.18}
\usepackage[section]{placeins}
\newcommand{\jpsi}{\ensuremath{J/\psi}}
\newcommand{\CP}{\ensuremath{CP}}
\newcommand{\Tr}{\operatorname{Tr}}
\newcommand{\gev}{\ensuremath{\mathrm{GeV}}}
\newcommand{\ecm}{\ensuremath{e\,\mathrm{cm}}}
\newcommand{\as}{\alpha_s}
\newcommand{\fas}{F_{\mathrm{as}}}
\title{\boldmath Timelike electric dipole form factors of the baryon octet: short-distance kernels and baryonic rescattering}
\author{Chao-Qiang Geng,}
\author{Xiang-Nan Jin}
\author{and Chia-Wei Liu}
\affiliation{School of Fundamental Physics and Mathematical Sciences,
Hangzhou Institute for Advanced Study, UCAS, Hangzhou 310024, China}
\emailAdd{cqgeng@ucas.ac.cn}
\emailAdd{xnjin@ucas.ac.cn}
\emailAdd{chiaweiliu@ucas.ac.cn}
\abstract{We calculate the leading-twist contribution of quark electric dipole
moments (EDMs) to the timelike form factors of the baryon octet at the
$J/\psi$ mass using lattice light-cone distribution amplitudes. For a
strange-quark EDM $d_s$, the short-distance $\Sigma/\Lambda$ ratio is
$-0.080^{+0.037}_{-0.098}$ over the lattice input ranges. This range lies
above the nonrelativistic quark-model value $-1/3$. In the meson-exchange
model, kaon exchange feeds the large $\Xi$ source into the $\Sigma$ channel
and makes the $\Sigma$ and $\Lambda$ form factors comparable in magnitude.
The short-distance $\Lambda$ form factor is about $5\times10^{-4}d_s$.
The BESIII measurement then gives $|d_s|<1.3\times10^{-15}\,e\,\mathrm{cm}$
at 95\% confidence level. With rescattering at a $1\,\mathrm{GeV}$ cutoff,
joint fits to projected $\Lambda$, $\Sigma^+$ and $\Xi$ sensitivities lower
the $\Lambda$-only upper limits by more than $25\%$ at BESIII and the Super
Tau-Charm Facility. Renormalization-scale variation changes the projected
limits by about a factor of two.}
\hypersetup{pdftitle={Timelike electric dipole form factors of the baryon octet: short-distance kernels and baryonic rescattering},pdfauthor={Chao-Qiang Geng, Xiang-Nan Jin, Chia-Wei Liu},pdfkeywords={hyperon electric dipole form factors, quark dipoles, baryon distribution amplitudes, perturbative QCD}}
\compress
\begin{document}
\maketitle

\section{Introduction}
\label{sec:intro}

Electric dipole moments (EDMs) probe flavor-diagonal \CP\ violation beyond
the Standard Model~\cite{Pospelov:2005pr,Chupp:2019}.
Hyperon EDMs probe the strange quark directly, as do anomaly-induced
asymmetries in $e^+e^-\to K^+K^-\pi^0$~\cite{Geng:2026zcq}.
A fixed-target experiment bounded the static $\Lambda$ EDM to
$|d_\Lambda|<1.5\times10^{-16}\,\ecm$~\cite{Pondrom:1981}.
BESIII uses angular correlations in entangled
$\jpsi\to\Lambda\bar\Lambda$ decays to constrain
$d_\Lambda$~\cite{BESIII:2025EDM}, following earlier proposals and studies
of the channel~\cite{BESIII:2018cnd,BESIII:2022qax,HeMaMcKellar:1993,Fu:2023,Du:2024}.
At 95\% confidence level (CL), the real part lies between $-8.6$ and $3.3$
in units of $10^{-19}\,\ecm$. The imaginary part lies between $-2.5$ and $7.2$
in the same units. The magnitude is bounded by $7.0\times10^{-19}\,\ecm$
at 95\% CL.

The decay probes the timelike EDM form factor at the charmonium mass.
It enters through the CP-violating form factor
$H_T=2eg_Vd_\Lambda/(3m_{J/\psi}^2)$, with $g_V$ fixed by the
$\jpsi\to e^+e^-$ width~\cite{BESIII:2025EDM,Du:2024}.
The same angular analysis applies to $\Sigma^+$ and
$\Xi$~\cite{BESIII:2020fqg,BESIII:2025jxt,BESIII:2021ypr,BESIII:2026hgj,BESIII:2023drj}.
The projected EDM sensitivities are $10^{-19}$--$10^{-18}\,\ecm$ at
BESIII and one to two orders of magnitude better at the Super Tau-Charm
Facility (STCF)~\cite{Achasov:2023gey,Cheng:2025kpp,Fu:2023,Guo:2025Polarized}.

At large momentum transfer, an exclusive baryon form factor factorizes into
a hard kernel convoluted with the light-cone distribution amplitudes
(LCDAs)~\cite{LepageBrodsky:1980}. Reference~\cite{Chen:2026} applied this
to the $\Lambda$ EDM form factor from quark EDMs and chromoelectric dipole
moments (CEDMs). At leading power the hard contribution scales as $1/s^2$,
where $s$ is the pair invariant mass squared~\cite{Brodsky:1973kr,Lepage:1979za,Chernyak:1983ej}.
Soft and higher-twist contributions can be important at a few
GeV~\cite{BraunLenzWittmann:2006}. Final-state interactions can also affect
the baryon--antibaryon form factors~\cite{HaidenbauerMeissner:2016,Haidenbauer:2021Hyperon,Chen:2006yn,Geng:2026hcm}.

We extend the leading-twist quark-EDM calculation to the whole octet.
The lattice $\Sigma$ and $\Xi$ LCDAs have the opposite exchange symmetry
to the $\Lambda$ LCDAs~\cite{Bali:2016Octet,RQCD:2019hps,Bali:2025Octet}.
The flavor multiplicity of the valence pair enters the hard-scattering coefficients.
QCD sum rules and large-momentum effective theory give independent access to
the LCDAs~\cite{Chernyak:1987nu,LiuHuang:2009SigmaLambda,LiuHuang:2009Xi,LatticeParton:2024vck,LPC:2026lcj,LPC:2026mvw}.
For a strange-quark EDM, the leading-twist $\Sigma/\Lambda$ coefficient ratio
is about $-0.080$. Its lattice-input range lies above the nonrelativistic
quark-model (NRQM) value $-1/3$.
We include rescattering of the strange-quark source through $\pi$, $K$ and
$\eta_8$ exchange in the isoscalar ${}^1P_1$ channel. We use the resulting
form factors to fit the strange-quark EDM $d_s$ jointly across hyperon channels.

The neutron EDM is sensitive to $d_s$ through the strange tensor
charge~\cite{Bhattacharya:2015esa}. Lattice QCD gives
$g_T^s=-0.0027(16)$~\cite{Gupta:2018lvp,FlavourLatticeAveragingGroupFLAG:2024oxs}.
At this central value the neutron limit $|d_n|<1.8\times10^{-26}\,\ecm$ at
90\% CL~\cite{Abel:2020pzs} would give $|d_s|\lesssim7\times10^{-24}\,\ecm$
if $d_s$ alone contributes. The 95\% CL range of $g_T^s$ includes zero and
leaves $d_s$ unconstrained at that level. The strange quark is a valence quark
of the hyperons. At $m_{J/\psi}$ the hard contribution gives
$d_\Lambda\simeq5\times10^{-4}\,d_s$, three orders of magnitude below the
static identification $d_\Lambda=d_s$.

We truncate the leading-twist LCDAs at first order in the conformal expansion
and hold the strange-EDM Born source fixed inside the rescattering loop.
The phase between gluonic and electromagnetic production is undetermined.
We therefore use the magnitudes of the form factors to estimate the joint
sensitivity. Section~\ref{sec:sd} gives the short-distance calculation and
its inputs, section~\ref{sec:baryon-fsi} gives the rescattering correction, and
section~\ref{sec:numerics} gives the numerical results and joint fit.
We conclude in section~\ref{sec:conclusions}.

\section{Short-distance contribution}
\label{sec:sd}

\label{sec:operators}
The quark EDM interactions are~\cite{Chen:2026}
\begin{equation}
 {\cal L}_{\rm EDM}=-\frac{i}{2}\sum_{q=u,d,s}
 d_q\bar q\sigma^{\mu\nu}\gamma_5qF_{\mu\nu}.
 \label{eq:operators}
\end{equation}
Our signs correspond to
$D_\mu=\partial_\mu+iee_qA_\mu-ig_sT^aG^a_\mu$, with $e_q$ the quark charge
in units of $e>0$. For baryon and antibaryon momenta $p$ and $\bar p$ we use
$s=(p+\bar p)^2$. In the $\gamma^*B\bar B$ vertex the timelike EDM form factor
$d_B(s)$ multiplies $\mathcal D^\mu=\sigma^{\mu\nu}(p+\bar p)_\nu\gamma_5$.
For quark EDMs we obtain
\begin{equation}
 d_B(s)=\sum_{q=u,d,s}\sum_{B'}
 \bigl(\delta_{BB'}+\widehat R^q_{BB'}(s)\bigr)
 h_{B'q}(s)d_q(\sqrt{s}).
 \label{eq:coefficients}
\end{equation}
The sum over $B'$ runs over the octet. The real, dimensionless coefficients
$h_{B'q}=O(\as^2)$ weight $d_q$ in the short-distance $B'\bar B'$ form factor.
They come from diagrams with two hard gluons connecting the three valence-quark
lines. The response $\widehat R^q_{BB'}$ sums rescattering from $B'\bar B'$ to
$B\bar B$ over the exchanged mesons for an EDM source of flavor $q$.

Low-energy matrix elements such as tensor charges govern the static EDM $d_B(0)$~\cite{Gupta:2018lvp,Bali:2023sdi}.
Quark CEDMs contribute to $d_B(s)$ and also enter charmonium decays
through three-gluon annihilation~\cite{Chen:2026}. The measured relative
phase of the $\Lambda$ electromagnetic form factors illustrates that timelike
form factors can be complex~\cite{BESIII:2019nep,BESIII:2018cnd,BESIII:2022qax}.

\label{sec:lcda}
We order the valence quarks as $q_1q_2\,q_3$, with $q_3$ carrying the baryon
spinor index. The diquark $q_1q_2$ is in square brackets for the flavor
antisymmetric case and in curly brackets for the symmetric case, giving $p=\{uu\}d$, $n=\{dd\}u$, $\Lambda=[ud]s$,
$\Sigma^{+,-}=\{uu\}s,\{dd\}s$ and $\Xi^{0,-}=\{ss\}u,\{ss\}d$.
The leading-twist LCDAs $V^B$, $A^B$ and $T^B$ are defined by the
three-quark matrix element along a lightlike vector
$z$~\cite{Braun:2000kw,Bali:2016Octet,Bali:2025Octet}
\begin{multline}
 4\langle0|\varepsilon^{ijk}q^i_{1\alpha}(a_1z)\,q^j_{2\beta}(a_2z)\,
 q^k_{3\gamma}(a_3z)|B(p)\rangle
 =\int[dx]\,e^{-ip\cdot z\sum_jx_ja_j}
 \bigl[V^B(x)(\slashed pC)_{\alpha\beta}(\gamma_5u_B)_\gamma\\
 +A^B(x)(\slashed p\gamma_5C)_{\alpha\beta}(u_B)_\gamma
 +T^B(x)(i\sigma_{\mu\nu}p^\nu C)_{\alpha\beta}
 (\gamma^\mu\gamma_5u_B)_\gamma\bigr].
 \label{eq:lcda-def}
\end{multline}
Here $z^2=0$, the $a_j$ are real, $C$ is the charge-conjugation matrix,
and Wilson lines are implied. Quark $q_i$ carries momentum fraction $x_i$
with $\int[dx]=\int dx_1dx_2dx_3\,\delta(1-x_1-x_2-x_3)$.
We use the three-quark operator renormalization prescription of
ref.~\cite{KranklManashov:2011}. At first order in the conformal expansion each
LCDA is a linear polynomial times the asymptotic factor
$\fas(x)=120x_1x_2x_3$~\cite{Braun:1999te,Bali:2016Octet}.

For $B\ne\Lambda$, $V$ and $T$ are symmetric and $A$ is antisymmetric under
$x_1\leftrightarrow x_2$. The antisymmetric $[ud]$ pair reverses these
symmetries for the $\Lambda$. Reference~\cite{Bali:2025Octet} uses the same
quark ordering. With $x_{ijk}\equiv(x_i,x_j,x_k)$ its definitions read
\begin{align}
 \Phi^{B\ne\Lambda}_\pm(x_{123})&=\tfrac12\bigl\{[V-A]^B(x_{123})
 \pm[V-A]^B(x_{321})\bigr\}
 \label{eq:bali-def-octet}\\
 \Phi^\Lambda_+(x_{123})&=\tfrac1{\sqrt6}\bigl\{[V-A]^\Lambda(x_{123})
 +[V-A]^\Lambda(x_{321})\bigr\}
 \label{eq:bali-def-lambda-plus}\\
 \Phi^\Lambda_-(x_{123})&=-\sqrt{\tfrac32}\bigl\{[V-A]^\Lambda(x_{123})
 -[V-A]^\Lambda(x_{321})\bigr\}
 \label{eq:bali-def}\\
 \Pi^{B\ne\Lambda}(x_{123})&=T^B(x_{132})\qquad
 \Pi^\Lambda(x_{123})=\sqrt6\,T^\Lambda(x_{132}).
 \label{eq:bali-tensor-def}
\end{align}
The first-order conformal expansion gives
\begin{align}
 \Phi^B_+&=\fas(x)\bigl[\varphi^B_{00}+7\varphi^B_{11}(x_1-2x_2+x_3)\bigr]
 &\Phi^B_-&=21\fas(x)\,\varphi^B_{10}(x_1-x_3)\label{eq:bali-even-odd}\\
 \Pi^{B\ne\Lambda}&=\fas(x)\bigl[\pi^B_{00}+7\pi^B_{11}(x_1-2x_2+x_3)\bigr]
 &\Pi^\Lambda&=21\fas(x)\,\pi^\Lambda_{10}(x_1-x_3).
 \label{eq:bali-moments}
\end{align}
The two projections give $[V-A]^B=\Phi^B_++\Phi^B_-$ for $B\ne\Lambda$
and $[V-A]^\Lambda=\sqrt{3/2}\,(\Phi^\Lambda_+-\Phi^\Lambda_-/3)$.
The $x_1\leftrightarrow x_2$ projections of $V-A$ with the parities of $V$
and $A$ then give $V$ and $-A$.
Equation~\eqref{eq:bali-tensor-def} gives $T^B(x_{123})=\Pi^B(x_{132})$ for
$B\ne\Lambda$ and $T^\Lambda(x_{123})=\Pi^\Lambda(x_{132})/\sqrt6$.

\label{sec:hard}
We evaluate the 42 EDM diagrams with color factor $2/27$~\cite{Chen:2026}.
Figure~\ref{fig:long-distance}(a) shows a representative diagram. We use
$\kappa=2(4\pi\as)^2/(27s^2)$. In the assignments above $q_3$ is distinct from
$q_1$ and $q_2$. Its short-distance coefficient is
\begin{equation}
 \begin{aligned}
 h_{Bq_3}&=\frac{\kappa}{1+\delta_{q_1q_2}}\int[dx]\,[dy]\,
 \Bigl\{\bigl[V^B(x)\bar V^B(y)-A^B(x)\bar A^B(y)\bigr]\mathcal H_s(x,y)\\
 &\hspace{6em}+\bigl[V^B(x)\bar A^B(y)-A^B(x)\bar V^B(y)\bigr]\mathcal C_s(x,y)\Bigr\}.
 \end{aligned}
 \label{eq:h-third-quark}
\end{equation}
For identical quarks the two fermion routings and the inverse Fock factors
on both sides give $2/(2!)^2=1/2$. The EDM can act on either quark in the
pair, giving a compensating factor of two. The pair coefficients are
\begin{equation}
 h_{Bq_1}=h_{Bq_2}=\kappa\int[dx]\,[dy]\,
 \bigl[V^B(x)+A^B(x)\bigr]\bar T^B(y)\,\mathcal H_q(x,y).
 \label{eq:h-quark-pair}
\end{equation}
Here $y_i$ is the momentum fraction of $\bar q_i$. Charge conjugation gives
the antibaryon LCDAs $\bar V^B$, $\bar A^B$ and $\bar T^B$.
The kernels $\mathcal H_s$, $\mathcal C_s$ and $\mathcal H_q$ of
ref.~\cite{Chen:2026} use the reverse quark ordering. We apply them after
$x_1\leftrightarrow x_3$ and $y_1\leftrightarrow y_3$.
That reference used the $\Lambda$ LCDA symmetry under the joint exchange
$x_1\leftrightarrow x_2$, $y_1\leftrightarrow y_2$ to simplify the kernels.
The other octet LCDAs reverse the parities of both factors in each product
in eqs.~\eqref{eq:h-third-quark} and~\eqref{eq:h-quark-pair}, preserving the
joint parity. A direct evaluation of the diagrams confirms that the kernels
apply to the whole octet. For the neutral $\Sigma$, isospin gives the average
of the $\Sigma^+$ and $\Sigma^-$ coefficients for each quark flavor.

For $q_1=q_2$ we insert the LCDAs reconstructed from eqs.~\eqref{eq:bali-even-odd}
and~\eqref{eq:bali-moments}. With $f^B=\varphi^B_{00}$ and $f_T^B=\pi^B_{00}$
and the baryon superscript suppressed, the coefficients are
\begin{align}
 h_{Bq_3}&=-25\kappa(3f-63\varphi_{10}+7\varphi_{11})(3f-49\varphi_{10}+21\varphi_{11}).
 \label{eq:h-third-quark-analytic}\\
 h_{Bq_1}&=100\kappa\bigl(9ff_T+42f\pi_{11}+168f_T\varphi_{10}+42f_T\varphi_{11} +833\varphi_{10}\pi_{11}+147\varphi_{11}\pi_{11}\bigr).
 \label{eq:h-quark-pair-analytic}
\end{align}
The strange quark occupies position $3$ in the $\Sigma$ and positions
$1$ and $2$ in the $\Xi$, so their strange-EDM coefficients depend on
different combinations of shape parameters.

Table~\ref{tab:inputs} gives the lattice normalizations and shape parameters
at $\mu_0=2\,\gev$~\cite{Bali:2025Octet}. For the central $J/\psi$ results we
use $\mu=m_{J/\psi}\simeq3.097\,\gev$ for the LCDAs, $\as$ and the dipoles.
The LCDAs obey the one-loop evolution law~\cite{LepageBrodsky:1980,Bali:2016Octet}
\begin{equation}
 \varphi_{nk}(\mu)=\varphi_{nk}(\mu_0)
 \left[\frac{\as(\mu)}{\as(\mu_0)}\right]^{\gamma_{nk}/\beta_0}
 \qquad (\gamma_{00},\gamma_{10},\gamma_{11})
 =\left(\frac23,\frac{26}{9},\frac{10}{3}\right).
 \label{eq:evolution}
\end{equation}
The $\pi_{nk}$ evolve in the same way, so $f=\varphi_{00}$ and
$f_T=\pi_{00}$ evolve with $\gamma_{00}$~\cite{Bali:2016Octet}. We
use $\as(\mu)=4\pi/[\beta_0\ln(\mu^2/\Lambda_{\rm QCD}^2)]$ with
$\beta_0=11-2n_f/3$, $n_f=4$ and $\Lambda_{\rm QCD}=0.2\,\gev$, which
gives $\as(m_{J/\psi})=0.275$ and $\as(2\,\gev)=0.327$. The three-flavor
lattice inputs are used without threshold or scheme conversions.
The reference value is $\as(m_Z)=0.118$~\cite{ParticleDataGroup:2026mpi}.
Three-loop running from this value gives $\as(m_{J/\psi})\simeq0.25$
and lowers the coefficients by about $18\%$, within the scale variation below.
Next-to-leading-order corrections have been computed for the nucleon and
$\Sigma$ form factors~\cite{Chen:2024fhj,Huang:2024ugd,Shi:2026nxz}.

To propagate the ranges in table~\ref{tab:inputs}, we preserve the isospin
and nucleon relations and add each parameter's statistical, renormalization,
continuum and chiral errors linearly~\cite{Bali:2025Octet}.
We treat the $1.2\%$ scale-setting error as a common rescaling of all
dimensionful inputs. The extrema of each $h_{Bs}$ lie at the corners of the
resulting parameter box.

\begin{table}[tb]
 \centering
 \renewcommand{\arraystretch}{1.45}
 \begin{tabular}{lrrrr}
 \toprule
 Parameter & $N$ & $\Sigma$ & $\Xi$ & $\Lambda$ \\
 \midrule
 $f$ & $3.29^{+0.34}_{-0.32}$ & $5.32^{+0.18}_{-0.18}$ & $6.15^{+0.38}_{-0.37}$ & $4.75^{+0.26}_{-0.24}$ \\
 $f_T$ & $3.29^{+0.34}_{-0.32}$ & $5.15^{+0.18}_{-0.17}$ & $6.34^{+0.40}_{-0.40}$ & --- \\
 $\varphi_{10}$ & $0.181^{+0.031}_{-0.027}$ & $0.094^{+0.019}_{-0.051}$ & $0.361^{+0.050}_{-0.061}$ & $0.563^{+0.109}_{-0.103}$ \\
 $\varphi_{11}$ & $0.118^{+0.041}_{-0.039}$ & $0.200^{+0.059}_{-0.061}$ & $-0.001^{+0.016}_{-0.015}$ & $0.242^{+0.071}_{-0.069}$ \\
 $\pi_{11}$ & $0.118^{+0.041}_{-0.039}$ & $-0.080^{+0.033}_{-0.033}$ & $0.399^{+0.104}_{-0.106}$ & --- \\
 $\pi_{10}$ & --- & --- & --- & $0.237^{+0.065}_{-0.070}$ \\
 \bottomrule
 \end{tabular}
 \space
 \caption{Lattice LCDA parameters at $\mu_0=2\,\gev$ in units of
 $10^{-3}\,\gev^2$~\cite{Bali:2025Octet}. The quoted errors sum the
 statistical, renormalization, continuum and chiral errors linearly.
 The common $1.2\%$ scale-setting uncertainty is treated separately.
 For the nucleon $f_T^N=f^N$ and
 $\pi_{11}^N=\varphi_{11}^N$. Missing entries vanish by symmetry, and
 for the $\Lambda$ the entry $f$ is $\varphi_{00}^\Lambda$.}
 \label{tab:inputs}
\end{table}

\begin{table}[tb]
 \centering\small
 \setlength{\tabcolsep}{8pt}
 \renewcommand{\arraystretch}{1.35}
 \begin{tabular}{lrrr}
 \toprule
 $B$ & $h_{Bu}$ & $h_{Bd}$ & $h_{Bs}$ \\
 \midrule
 $p$ & $2.26^{+0.77}_{-0.61}$ & $0.002^{+0.013}_{-0.046}$ & $0$ \\
 $n$ & $0.002^{+0.013}_{-0.046}$ & $2.26^{+0.77}_{-0.61}$ & $0$ \\
 $\Lambda$ & $0.462^{+0.418}_{-0.291}$ & $0.462^{+0.418}_{-0.291}$ & $5.31^{+1.67}_{-1.35}$ \\
 $\Sigma^+$ & $3.19^{+0.64}_{-0.78}$ & $0$ & $-0.423^{+0.142}_{-0.313}$ \\
 $\Sigma^-$ & $0$ & $3.19^{+0.64}_{-0.78}$ & $-0.423^{+0.142}_{-0.313}$ \\
 $\Xi^0$ & $0.012^{+0.004}_{-0.051}$ & $0$ & $8.65^{+2.20}_{-2.03}$ \\
 $\Xi^-$ & $0$ & $0.012^{+0.004}_{-0.051}$ & $8.65^{+2.20}_{-2.03}$ \\
 \bottomrule
 \end{tabular}
 \space
 \caption{EDM coefficients $h_{Bq}$ of eq.~\eqref{eq:coefficients} at
 $m_{J/\psi}$, in units of $10^{-4}$. The errors come from the input
 ranges of table~\ref{tab:inputs}, including the common $1.2\%$ scale-setting
 uncertainty.}
 \label{tab:coefficients}
\end{table}

\begin{table}[tb]
 \centering\small
 \begin{tabular}{lrrrrr}
 \toprule
 $\mu$ & $10^4h_{\Lambda s}$ & $10^4h_{\Sigma s}$ & $10^4h_{\Xi s}$ & $h_{\Sigma s}/h_{\Lambda s}$ & $h_{\Xi s}/h_{\Lambda s}$ \\
 \midrule
 $m_{J/\psi}/2$ & $10.8$ & $-0.774$ & $17.3$ & $-0.0717$ & $1.60$ \\
 $m_{J/\psi}$ & $5.31$ & $-0.423$ & $8.65$ & $-0.0797$ & $1.63$ \\
 $2m_{J/\psi}$ & $3.07$ & $-0.265$ & $5.07$ & $-0.0862$ & $1.65$ \\
 \midrule
 NRQM, $s=0$ & $10^4$ & $-10^4/3$ & $4\times10^4/3$ & $-1/3$ & $4/3$ \\
 \bottomrule
 \end{tabular}
 \space
 \caption{Scale dependence of the strange-EDM coefficients and of their
 ratios, compared with the NRQM at $s=0$.}
 \label{tab:scale}
\end{table}

Table~\ref{tab:coefficients} gives the coefficients at $m_{J/\psi}$.
With the same lattice inputs ref.~\cite{Chen:2026} found $10^4 h_{\Lambda s}=5.29$
and $10^4 h_{\Lambda u}=10^4 h_{\Lambda d}=0.461$. Our coefficients agree
within $0.5\%$. The small entries $h_{pd}$ and $h_{nu}$ arise from a
near cancellation in eq.~\eqref{eq:h-third-quark-analytic}. Their signs
change under scale variation and when the nucleon $\varphi_{10}$ is lowered
by its statistical error.

We vary the common renormalization and factorization scale over
$[m_{J/\psi}/2,2m_{J/\psi}]$, evolving the LCDAs and $\as$ while holding the
dipoles fixed at $m_{J/\psi}$. Lowering the scale to $m_{J/\psi}/2$ roughly
doubles the strange coefficients. Raising it to $2m_{J/\psi}$ reduces them
by about $40\%$. The ratios vary little because the leading power of $\as$
cancels (table~\ref{tab:scale}). The scale-setting uncertainty changes the
coefficients by $2.4\%$.

\label{sec:su6}

The short-distance coefficients obey
$|h_{\Xi s}|>|h_{\Lambda s}|\gg|h_{\Sigma s}|$. Varying the lattice inputs
in table~\ref{tab:inputs} independently for each baryon gives
\begin{equation}
 \frac{h_{\Sigma s}}{h_{\Lambda s}}=-0.080^{+0.037}_{-0.098}\qquad
 \frac{h_{\Xi s}}{h_{\Lambda s}}=1.63^{+0.99}_{-0.64}.
 \label{eq:ratios}
\end{equation}
The NRQM ratios are recovered with vanishing shape parameters and a
common normalization $f^B=f_T^B=\varphi_{00}^\Lambda=f$. In this limit
the coefficients are $h_{\Sigma s}=-225\kappa f^2$ and $h_{\Xi s}=900\kappa f^2$.
The $\Lambda$ kernels give $h_{\Lambda s}=675\kappa f^2$~\cite{Chen:2026}.
These values give the NRQM ratios~\cite{Du:2024}
\begin{equation}
 \frac{h_{\Sigma s}}{h_{\Lambda s}}=-\frac13\qquad
 \frac{h_{\Xi s}}{h_{\Lambda s}}=\frac43.
 \label{eq:su6-limit}
\end{equation}
SU(3) symmetry alone does not fix these ratios. With the lattice
normalizations and vanishing shape parameters we obtain
$(h_{\Lambda s},h_{\Sigma s},h_{\Xi s})=(1.43,-0.596,3.29)\times10^{-4}$.
Restoring the first-order terms enhances the $\Lambda$ and $\Xi$
coefficients by factors of $3.7$ and $2.6$. It reduces the magnitude of the
$\Sigma$ coefficient by $30\%$. The local-operator lattice inputs stop at
first order in the conformal expansion. Calculations of the full $x$
dependence are available for the $\Lambda$~\cite{LatticeParton:2024vck,LPC:2026lcj,LPC:2026mvw}.

\section{Baryonic long-distance contribution}
\label{sec:baryon-fsi}

In charmonium decays the dipole form factor enters through
$\jpsi\to\gamma^*\to B\bar B$~\cite{HeMaMcKellar:1993,Chen:2026}, and the dominant
CP-even amplitude proceeds through gluons~\cite{BrodskyLepage:1981,Chernyak:1983ej,Bolz:1997as,Kivel:2022fzk}.
The dipole source produces a spin-singlet $P$ wave. We focus on the strange-quark
source $q=s$ in eq.~\eqref{eq:coefficients} and calculate $\widehat R^s_{BB'}$ from single-meson
exchange between the baryon and antibaryon~\cite{Chen:2006yn}. For each
exchanged meson $M$ we evaluate the triangle loop including its two-baryon cut.

The real Born source is $d_{B,\mathrm{in}}=h_{Bs}d_s(\sqrt{s})$. The form factor per
unit strange-quark EDM is $r_B\equiv d_B/d_s$. At this order $h_{ps}=h_{ns}=0$, so only
hyperon Born sources contribute to the single-exchange correction to $d_B$.
For the proton, the pairs $(B',M)$ are $(\Lambda,K^+)$, $(\Sigma^0,K^+)$ and
$(\Sigma^+,K^0)$. For the $\Sigma^+$ they are $(\Sigma^+,\pi^0)$,
$(\Sigma^0,\pi^+)$, $(\Lambda,\pi^+)$, $(\Sigma^+,\eta_8)$ and $(\Xi^0,K^+)$.
For the $\Xi^0$ the pairs are $(\Xi^0,\pi^0)$, $(\Xi^-,\pi^+)$, $(\Xi^0,\eta_8)$,
$(\Lambda,\bar K^0)$, $(\Sigma^0,\bar K^0)$ and $(\Sigma^+,K^-)$.
Isospin symmetry relates the neutron, $\Sigma^{0,-}$ and $\Xi^-$ to their partners.
The restriction to hyperon intermediate states also applies to the $\Lambda$.

We take the meson--baryon interaction from the octet axial coupling~\cite{Polinder:2006}
\begin{equation}
 \mathcal L_{BB\pi}=\frac1{2f_0}\Tr\!\left[
 \bar{\mathcal B}\gamma^\alpha\gamma_5
 \bigl(D\{\partial_\alpha\pi,\mathcal B\}
       +F[\partial_\alpha\pi,\mathcal B]\bigr)\right].
 \label{eq:fsi-axial}
\end{equation}
Here $\mathcal B$ and $\pi$ are the baryon and pseudoscalar meson octet matrices
\begin{equation}
 \mathcal B=\begin{pmatrix}
  \frac{\Sigma^0}{\sqrt2}+\frac{\Lambda}{\sqrt6} & \Sigma^+ & p\\[2pt]
  \Sigma^- & -\frac{\Sigma^0}{\sqrt2}+\frac{\Lambda}{\sqrt6} & n\\[2pt]
  \Xi^- & \Xi^0 & -\frac{2\Lambda}{\sqrt6}
 \end{pmatrix}
 \qquad
 \pi=\begin{pmatrix}
  \pi^0+\frac{\eta_8}{\sqrt3} & \sqrt2\,\pi^+ & \sqrt2\,K^+\\[2pt]
  \sqrt2\,\pi^- & -\pi^0+\frac{\eta_8}{\sqrt3} & \sqrt2\,K^0\\[2pt]
  \sqrt2\,K^- & \sqrt2\,\bar K^0 & -\frac{2\eta_8}{\sqrt3}
 \end{pmatrix}.
 \label{eq:octet-matrices}
\end{equation}
For an exchanged pseudoscalar $M$, the term
$(c_{aiM}/f_0)\bar B_a\gamma^\alpha\gamma_5 B_i\partial_\alpha M$ in $\mathcal L$
defines $c_{aiM}$, including the factor of $1/2$ in eq.~\eqref{eq:fsi-axial}.
We take $D=0.80$, $F=0.46$~\cite{Cabibbo:2003cu} and $f_0=0.0922\,\gev$.
We sum over the intermediate charge states using the isospin-averaged masses~\cite{ParticleDataGroup:2026mpi}
\begin{align}
 (m_N,m_\Lambda,m_\Sigma,m_\Xi)&=(0.939,1.116,1.193,1.318)\,\gev
 \label{eq:fsi-baryon-masses}\\
 (m_\pi,m_K,m_{\eta_8})&=(0.138,0.496,0.567)\,\gev.
 \label{eq:fsi-masses}
\end{align}
The $\eta_8$ mass follows from the Gell-Mann--Okubo relation $m_{\eta_8}^2=(4m_K^2-m_\pi^2)/3$.
 
\begin{figure}[tb]
 \centering
 \begin{tikzpicture}[
   x=1cm,y=1cm,font=\small,
   line cap=round,line join=round,
   ld fermion/.style={draw=black,line width=.65pt,
     postaction={decorate},decoration={markings,
       mark=at position .56 with {\arrow{Latex[length=1.6mm,width=1.05mm]}}}},
   ld antifermion/.style={draw=black,line width=.65pt,
     postaction={decorate},decoration={markings,
       mark=at position .56 with {\arrowreversed{Latex[length=1.6mm,width=1.05mm]}}}},
   ld photon/.style={draw=black,line width=.65pt,decorate,
     decoration={snake,amplitude=1.15pt,segment length=5.2pt,
       pre length=1pt,post length=1pt}},
   sd gluon/.style={draw=black,line width=.65pt,decorate,
     decoration={coil,amplitude=1.5pt,segment length=3.4pt,aspect=.65,
       pre length=1pt,post length=1pt}},
   sd lcda/.style={draw=black,fill=black!8,line width=.65pt},
   ld meson/.style={draw=black,line width=.65pt,dash pattern=on 3pt off 2pt}
 ]
   \coordinate (dsource) at (1.52,0);
   \coordinate (gluonOneLeft) at (1.52,.36);
   \coordinate (gluonOneRight) at (2.60,.36);
   \coordinate (gluonTwoLeft) at (2.60,-.36);
   \coordinate (gluonTwoRight) at (3.68,-.36);
   \foreach \x in {1.52,2.60,3.68} {
     \draw[line width=.65pt] (\x,-.50) -- (\x,.50);
   }
   \foreach \side/\valenceFrom/\valenceTo/\fermion in {
     1/-.52/-.12/ld fermion,
     -1/-.12/-.52/ld antifermion%
   } {
     \begin{scope}[yscale=\side,line width=.65pt]
       \begin{scope}[shift={(4.51,1.17)},rotate=14]
         \foreach \leg/\y in {outer/.24,middle/0,inner/-.24} {
           \coordinate (\leg-tail) at (-.52,\y);
           \draw[ld fermion] (\valenceFrom,\y) -- (\valenceTo,\y);
         }
         \draw[sd lcda] (-.12,-.36) rectangle (.12,.36);
         \draw[\fermion] (.12,0) -- (1.46,0);
       \end{scope}
       \draw (1.52,.50) .. controls (1.52,.94) and (3.35,1.14) .. (outer-tail);
       \draw (2.60,.50) .. controls (2.60,.80) and (3.59,.95) .. (middle-tail);
       \draw (3.68,.50) .. controls (3.68,.69) and (3.89,.77) .. (inner-tail);
     \end{scope}
   }
   \node[above=3pt] at (5.30,1.28) {$B(p)$};
   \node[below=3pt] at (5.30,-1.28) {$\bar B(\bar p)$};
   \draw[ld photon] (0,0) -- (dsource);
   \node[above=4pt] at (.64,0) {$\gamma^*$};
   \draw[sd gluon] (gluonOneLeft) -- (gluonOneRight);
   \draw[sd gluon] (gluonTwoLeft) -- (gluonTwoRight);
   \node[above=1.5pt] at (2.06,.36) {$g$};
   \node[below=1.5pt] at (3.14,-.36) {$g$};
   \foreach \vertex in {gluonOneLeft,gluonOneRight,gluonTwoLeft,gluonTwoRight} {
     \fill (\vertex) circle (1.15pt);
   }
   \filldraw[fill=white,draw=black,line width=.65pt] (dsource) circle (.145);
   \draw[line width=.65pt] (1.465,-.055) -- (1.575,.055);
   \draw[line width=.65pt] (1.465,.055) -- (1.575,-.055);
   \node[anchor=north east] at (1.39,-.12) {$d_q$};
   \node at (2.98,-2.20) {(a) Short-distance production};

   \coordinate (rsource) at (7.96,0);
   \coordinate (upper) at (10.25,1.02);
   \coordinate (lower) at (10.25,-1.02);
   \draw[ld photon] (6.46,0) -- (rsource);
   \node[above=4pt] at (7.10,0) {$\gamma^*$};
   \draw[ld fermion] (rsource) -- (upper);
   \draw[ld antifermion] (rsource) -- (lower);
   \node[above=3pt] at (9.16,.55) {$B'$};
   \node[below=3pt] at (9.16,-.55) {$\bar B'$};
   \draw[ld meson] (upper) -- (lower);
   \node[right=4pt] at (10.25,0) {$M$};
   \draw[ld fermion] (upper) -- (12.70,1.36);
   \draw[ld antifermion] (lower) -- (12.70,-1.36);
   \node[above=3pt] at (11.75,1.23) {$B(p)$};
   \node[below=3pt] at (11.75,-1.23) {$\bar B(\bar p)$};
   \fill (upper) circle (1.35pt);
   \fill (lower) circle (1.35pt);
   \filldraw[fill=white,draw=black,line width=.65pt] (rsource) circle (.145);
   \draw[line width=.65pt] (7.905,-.055) -- (8.015,.055);
   \draw[line width=.65pt] (7.905,.055) -- (8.015,-.055);
   \node[below=10pt] at (rsource) {$d_{B',\mathrm{in}}$};
   \node at (9.56,-2.20) {(b) One-meson rescattering};
 \end{tikzpicture}
 \space
 \caption{(a) A representative short-distance diagram with a quark EDM
 insertion $d_q$ and two hard gluons. The boxes denote the baryon and
 antibaryon LCDA projections. (b) Rescattering of the Born source
 $h_{B's}d_s$ through one-meson exchange with $M=\pi,K,\eta_8$.}
 \label{fig:long-distance}
\end{figure}
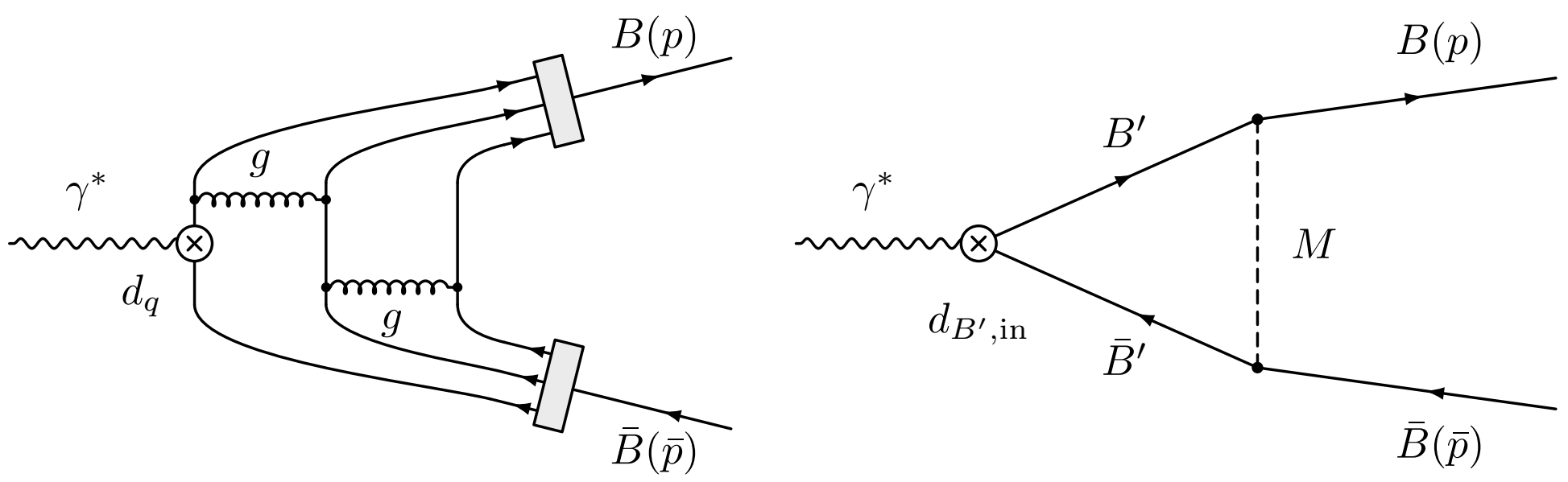

We project the on-shell vertex $i\Gamma^\mu$ between $\bar u(p)$ and
$v(\bar p)$ onto $\mathcal D^\mu$. For external mass $m$ the coefficient is
\begin{equation}
 d=\Tr\bigl[\mathbb P^D_\mu\Gamma^\mu\bigr]\qquad
 \mathbb P^D_\mu=
 \frac{[g_{\mu\rho}-(p+\bar p)_\mu(p+\bar p)_\rho/s]\,(\slashed{\bar p}-m)\,
 \sigma^{\rho\nu}(p+\bar p)_\nu\gamma_5\,(\slashed p+m)}{2s(s-4m^2)}.
 \label{eq:edm-projector}
\end{equation}
We hold the Born source fixed inside the loop and attach a monopole factor
from meson-exchange models to each strong vertex~\cite{Machleidt:1987hj,Haidenbauer:2005zh}
\begin{equation}
 F_M(\ell^2)=\frac{\Lambda_{\text{cut-off}}^2-m_M^2}
                         {\Lambda_{\text{cut-off}}^2-\ell^2-i0}\qquad
 \Lambda_{\text{cut-off}}=1\,\gev.
 \label{eq:fsi-strong-ff}
\end{equation}
We also evaluate the loop at $\Lambda_{\text{cut-off}}=0.9$ and $1.2\,\gev$.
With the baryon propagator $S_{B'}(k)=(\slashed k+m_{B'})/(k^2-m_{B'}^2+i0)$,
the diagram in figure~\ref{fig:long-distance}(b) gives
\begin{equation}
 \begin{aligned}
 \widehat R^s_{BB'}(s)&=
 -\frac1{f_0^2}\sum_M c_{BB'M}c_{B'B\bar M}\\
 &\quad\times\int\frac{d^4\ell}{i(2\pi)^4}\,
 \frac{F_M^2(\ell^2)}{\ell^2-m_M^2+i0}\,
 \Tr\!\left[
 \mathbb P^D_\mu\slashed\ell\gamma_5 S_{B'}(p+\ell)
 \mathcal D^\mu S_{B'}(\ell-\bar p)\slashed\ell\gamma_5
 \right].
 \end{aligned}
 \label{eq:fsi-loop}
\end{equation}
The sum includes mesons with $c_{BB'M}c_{B'B\bar M}\ne0$, where $\bar M$ is
the conjugate of $M$. For example, eq.~\eqref{eq:fsi-axial} gives
$c_{p\Lambda K^+}=-(D+3F)/(2\sqrt3)$ for the proton's $(\Lambda,K^+)$ channel
and $c_{p\Lambda K^+}c_{\Lambda pK^-}/f_0^2=(D+3F)^2/(12f_0^2)$.

At $s=m_{J/\psi}^2$ and $\Lambda_{\text{cut-off}}=1\,\gev$, we sum
$\widehat R^s_{BB'}$ over the charge states of each intermediate multiplet.
The response in the order $B,B'=N,\Lambda,\Sigma,\Xi$ is
\begin{equation}
 \setlength{\arraycolsep}{3pt}
 1+\widehat R^s=\begin{pmatrix}
  0.579-0.553i & -0.082-0.068i & -0.019-0.015i & 0\\
  -0.231-0.206i & 0.954-0.032i & -0.325-0.397i & -0.017-0.010i\\
  -0.018-0.020i & -0.114-0.170i & 0.701-0.333i & -0.257-0.145i\\
  0 & -0.011-0.011i & -0.507-0.389i & 0.799-0.122i
 \end{pmatrix}.
 \label{eq:response}
\end{equation}
Isospin partners share these values. The $\Xi$ source feeds the $\Sigma$ through
$\widehat R^s_{\Sigma\Xi}=-0.257-0.145i$. This gives
$\widehat R^s_{\Sigma\Xi}h_{\Xi s}=(-2.22-1.25i)\times10^{-4}$. Its real part is about
five times the direct $\Sigma$ coefficient $h_{\Sigma s}=-0.423\times10^{-4}$.
The $\Xi$ contribution dominates the correction to the $\Sigma$.

We test higher iterations by replacing $1+\widehat R^s$ in eq.~\eqref{eq:response}
with the Omn\`es-type exponential $\exp(\widehat R^s)$, which has the same
first-order term. This shifts $|r_\Sigma|$ by $-17\%$, $|r_\Lambda|$ by $+4\%$ and
$|r_\Xi|$ by $+6\%$. The joint $d_s$ bounds of section~\ref{sec:numerics} stay
within $1\%$ of their first-iteration values. We use the first iteration for
the fit. For a static benchmark of the same loop model, we take $s\to0$ after extracting
the coefficient of the dipole tensor at $\Lambda_{\text{cut-off}}=1\,\gev$.
The NRQM source ratios are $(0,1,-1/3,4/3)$ in the order $(N,\Lambda,\Sigma,\Xi)$.
Rescattering maps them to $(-0.066,1.04,-0.575,1.29)$.

\section{Numerical results and joint fit}
\label{sec:numerics}

\begin{figure}[tb]
 \centering
 \begin{tikzpicture}[font=\small]
 \definecolor{ratioN}{RGB}{120,75,165}
 \definecolor{ratioLambda}{RGB}{0,114,178}
 \definecolor{ratioSigma}{RGB}{213,94,0}
 \definecolor{ratioXi}{RGB}{0,158,115}
 \begin{axis}[name=timelike,width=.92\textwidth,height=6.5cm,legend columns=2,legend cell align=left,legend style={at={(0.98,0.03)},anchor=south east,font=\footnotesize,draw=black!30,column sep=8pt,row sep=1pt},xmin=-5,xmax=10,ymin=-3.2,ymax=0.6,xlabel={$10^4\,\mathrm{Re}(r_B)$},ylabel={$10^4\,\mathrm{Im}(r_B)$},scaled ticks=false,axis lines=left,axis line style={black!65,line width=.55pt,-},tick style={black!65},tick align=outside,tick label style={font=\small},label style={font=\small},clip=false]
 \draw[black!30,densely dashed] (axis cs:-5,0)--(axis cs:10,0);
 \draw[black!30,densely dashed] (axis cs:0,-3.2)--(axis cs:0,0.6);
 \addplot[forget plot,domain=0:360,samples=241,draw=ratioN,fill=ratioN,fill opacity=.15,line width=.75pt] ({-0.42718997965214+(-0.17301582762972)*cos(x)+(0.00015420355897873)*sin(x)},{-0.35589641703849+(-0.14405142471216)*cos(x)+(-0.0001852092503317)*sin(x)}) -- cycle;
 \addplot[only marks,mark=*,mark size=3pt,color=ratioN,mark options={fill=ratioN,draw=white,line width=.6pt}] coordinates {(-0.427189979652,-0.355896417038)};
 \addlegendentry{$p,n$}
 \addplot[forget plot,domain=0:360,samples=241,draw=ratioLambda,fill=ratioLambda,fill opacity=.15,line width=.75pt] ({5.054627104113+(-2.0803850125666)*cos(x)+(-0.0023818000503582)*sin(x)},{-0.088396731998422+(0.025728199583632)*cos(x)+(-0.19259261074949)*sin(x)}) -- cycle;
 \addplot[only marks,mark=square*,mark size=3pt,color=ratioLambda,mark options={fill=ratioLambda,draw=white,line width=.6pt}] coordinates {(5.05462710411,-0.0883967319984)};
 \addlegendentry{$\Lambda$}
 \addplot[forget plot,domain=0:360,samples=241,draw=ratioSigma,fill=ratioSigma,fill opacity=.15,line width=.75pt] ({-3.1241982440969+(-0.95690371816264)*cos(x)+(0.19137182102789)*sin(x)},{-2.0126724062396+(-0.5728048080537)*cos(x)+(-0.31969774785126)*sin(x)}) -- cycle;
 \addplot[only marks,mark=triangle*,mark size=3pt,color=ratioSigma,mark options={fill=ratioSigma,draw=white,line width=.6pt}] coordinates {(-3.1241982441,-2.01267240624)};
 \addlegendentry{$\Sigma$}
 \addplot[forget plot,domain=0:360,samples=241,draw=ratioXi,fill=ratioXi,fill opacity=.15,line width=.75pt] ({7.0636943796188+(-2.4248801472637)*cos(x)+(-0.028982162617566)*sin(x)},{-0.94974425716544+(0.33369419180695)*cos(x)+(-0.21060681450747)*sin(x)}) -- cycle;
 \addplot[only marks,mark=diamond*,mark size=3pt,color=ratioXi,mark options={fill=ratioXi,draw=white,line width=.6pt}] coordinates {(7.06369437962,-0.949744257165)};
 \addlegendentry{$\Xi^0,\Xi^-$}
 \end{axis}
 \end{tikzpicture}
 \space
 \caption{Complex strange-EDM ratios at
 $m_{J/\psi}$ and $\Lambda_{\text{cut-off}}=1\,\gev$. Points give the central
 values, and the filled ellipses enclose the LCDA ranges at $\mu=m_{J/\psi}$,
 including the common lattice scale uncertainty.}
 \label{fig:complex-ratios}
\end{figure}
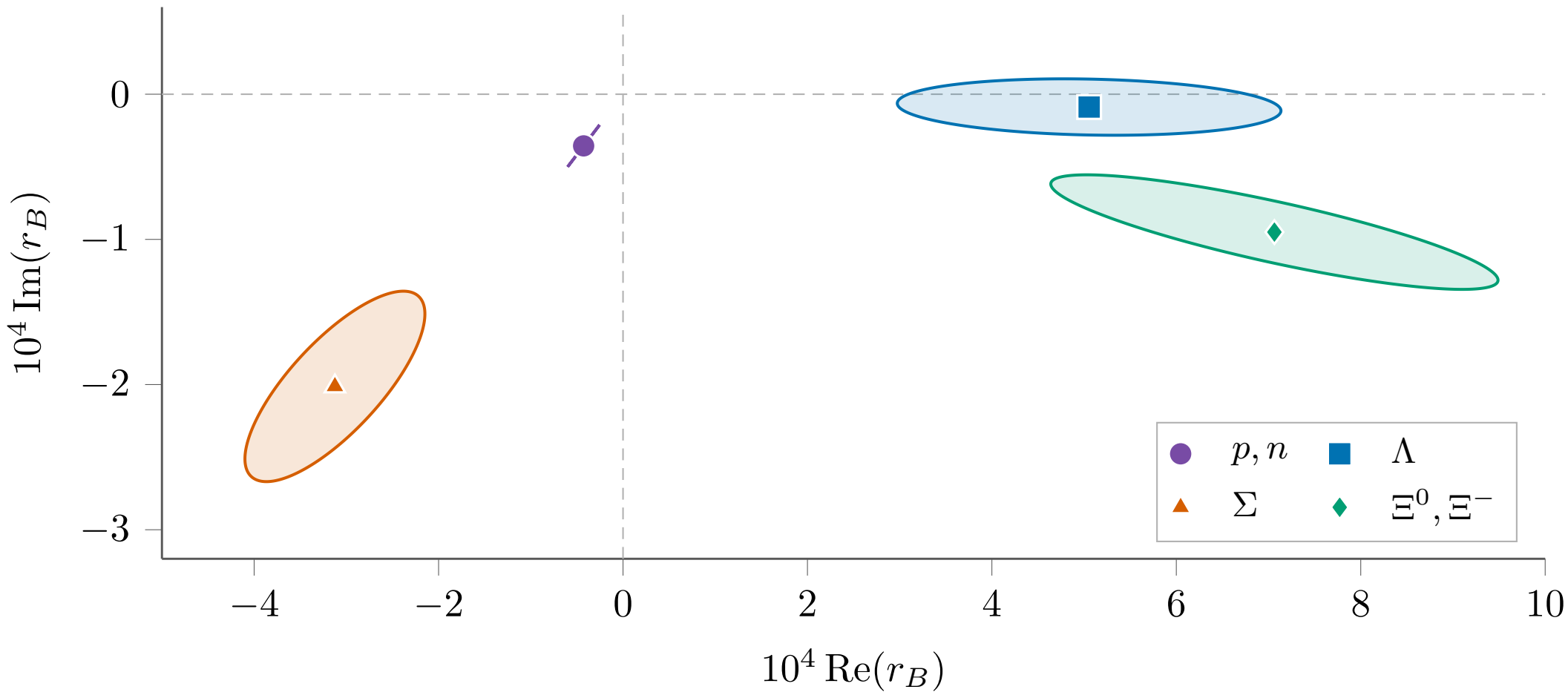

Table~\ref{tab:sd-ld-summary} gives the form-factor ratios $r_B=d_B/d_s$ at
$m_{J/\psi}$ for cutoffs of $0.9$, $1$ and $1.2\,\gev$. At the central cutoff
$\Lambda_{\text{cut-off}}=1\,\gev$, we estimate the scale uncertainty from
$\mu=m_{J/\psi}/2$ and $\mu=2m_{J/\psi}$ with central LCDA inputs. We hold
$s=m_{J/\psi}^2$, $d_s(m_{J/\psi})$ and $\widehat R^s$ fixed, so the shifts
come from the short-distance coefficients $h_{Bs}$. Table~\ref{tab:coefficients}
gives the central short-distance values. Figure~\ref{fig:complex-ratios} shows the LCDA input ranges in the complex
plane at $\mu=m_{J/\psi}$.

\begin{table}[tb]
 \centering
 \renewcommand{\arraystretch}{1.5}
 \setlength{\tabcolsep}{5pt}
 \begin{tabular}{lrrrrrr}
 \toprule
 & \multicolumn{2}{c}{$\Lambda_{\text{cut-off}}=0.9\,\gev$} & \multicolumn{2}{c}{$\Lambda_{\text{cut-off}}=1\,\gev$}
 & \multicolumn{2}{c}{$\Lambda_{\text{cut-off}}=1.2\,\gev$}\\
 \cmidrule(lr){2-3}\cmidrule(lr){4-5}\cmidrule(l){6-7}
 $B$ & $\mathrm{Re}\,r_B$ & $\mathrm{Im}\,r_B$ & $\mathrm{Re}\,r_B$ & $\mathrm{Im}\,r_B$
 & $\mathrm{Re}\,r_B$ & $\mathrm{Im}\,r_B$\\
 \midrule
 $N$ & $-0.265$ & $-0.267$ & $-0.427^{+0.180}_{-0.444}$ & $-0.356^{+0.150}_{-0.369}$ & $-0.865$ & $-0.513$\\
 $\Lambda$ & $5.17$ & $-0.0352$ & $5.05^{+5.21}_{-2.12}$ & $-0.088^{+0.045}_{-0.123}$ & $4.74$ & $-0.181$\\
 $\Sigma$ & $-2.24$ & $-1.67$ & $-3.12^{+1.28}_{-3.09}$ & $-2.01^{+0.84}_{-2.07}$ & $-5.27$ & $-2.51$\\
 $\Xi$ & $7.56$ & $-0.819$ & $7.06^{+7.01}_{-2.91}$ & $-0.95^{+0.40}_{-0.98}$ & $5.88$ & $-1.09$\\
 \bottomrule
 \end{tabular}
 \space
 \caption{Strange-EDM ratios at $m_{J/\psi}$ in units of $10^{-4}$ for
 three cutoffs. At $\Lambda_{\text{cut-off}}=1\,\gev$ the errors give the shifts for
 $\mu=m_{J/\psi}/2$ and $2m_{J/\psi}$. Isospin partners share the
 same values.}
 \label{tab:sd-ld-summary}
\end{table}

At $\Lambda_{\text{cut-off}}=1\,\gev$, rescattering reduces $|r_\Lambda|$ by
about $5\%$ and $|r_\Xi|$ by about $20\%$. The direct $\Sigma$ source is small.
Kaon exchange from the $\Xi$ accounts for about $80\%$ of the real $\Sigma$
shift and makes $|r_\Sigma|$ comparable to $|r_\Lambda|$. The nucleon has no
strange Born source at this order and receives its form factor almost
entirely from rescattering of the $\Lambda$ source.

The two-baryon cut retains the product $F_M^2(\ell^2)$ of the two meson
vertex factors in eq.~\eqref{eq:fsi-loop}, so the cutoff affects both parts
of the complex form factor. Over the cutoff range $0.9$ to $1.2\,\gev$,
$|\mathrm{Re}\,r_\Sigma|$ more than doubles. In the same range
$10^4\,\mathrm{Im}\,r_\Lambda$ changes from $-0.0352$ to $-0.181$.
The real parts for the $\Lambda$ and $\Xi$ decrease by about $8\%$ and
  $22\%$, respectively. Their own sources dominate both ratios. The magnitude
ratio $|d_\Xi|/|d_\Lambda|$ decreases from $1.47$ to $1.26$.

\label{sec:fit}

The BESIII measurement and the proposed analyses extract a complex $d_B$
from $H_T$~\cite{HeMa:2023,Du:2024,BESIII:2025EDM,Guo:2025Polarized}.
We identify this quantity with the timelike form factor
$d_B(m_{J/\psi}^2)$~\cite{Chen:2026}. The measured phase is relative to the
predominantly gluonic CP-even amplitude. Comparing it with the phase of
our form factor requires the relative phase of gluonic and electromagnetic
production~\cite{BaldiniFerroli:2019abd,Kivel:2022fzk}. We use the form-factor
magnitudes in the projections.

We project the sensitivities for null measurements with independent Gaussian
errors $\sigma_{d_B}$ on $|d_B|=|r_B|\,|d_s|$. Adding the inverse variances
for the individual channels gives
\begin{equation}
 \sigma_{d_s}=\Bigl(\sum_B\frac{|r_B|^2}{\sigma_{d_B}^2}\Bigr)^{-1/2}
 \qquad |d_s|<1.96\,\sigma_{d_s}\quad (95\%~\mathrm{CL}).
 \label{eq:fit-model}
\end{equation}
A full analysis would use the measured likelihoods for
$\mathrm{Re}\,d_B$ and $\mathrm{Im}\,d_B$.

\begin{table}[tb]
 \centering
 \renewcommand{\arraystretch}{1.15}
 \begin{tabular}{lrrrrrr}
 \toprule
  & \multicolumn{3}{c}{BESIII} & \multicolumn{3}{c}{STCF} \\
 \cmidrule(lr){2-4}\cmidrule(l){5-7}
 $B$ & $\sigma_{d_B}$ & $\sigma_{d_s}^{\rm SD}$ & $\sigma_{d_s}^{\rm full}$
     & $\sigma_{d_B}$ & $\sigma_{d_s}^{\rm SD}$ & $\sigma_{d_s}^{\rm full}$ \\
 \midrule
 $\Lambda$ & $36$ & $6.73$ & $7.06$ & $4.6$ & $0.867$ & $0.910$ \\
 $\Sigma^+$ & $41$ & $96.3$ & $11.0$ & $5.3$ & $12.5$ & $1.43$ \\
 $\Xi^-$ & $92$ & $10.7$ & $13.0$ & $12$ & $1.39$ & $1.68$ \\
 $\Xi^0$ & $126$ & $14.6$ & $17.7$ & $17$ & $1.97$ & $2.39$ \\
 \midrule
 combined &  & $5.30$ & $5.16$ &  & $0.687$ & $0.670$ \\
 \bottomrule
 \end{tabular}
 \space
 \caption{Projected standard errors for null measurements in units of
 $10^{-20}\,\ecm$ for $|d_B|$ and $10^{-16}\,\ecm$ for $|d_s(m_{J/\psi})|$.
 We use the unpolarized STCF projections and BESIII relative precisions of
 ref.~\cite{Guo:2025Polarized}, normalizing $1.96\,\sigma_{d_\Lambda}$ to the
 measured BESIII bound. The SD columns set $\widehat R^s=0$, and the full
 columns include rescattering at $\Lambda_{\text{cut-off}}=1\,\gev$.}
 \label{tab:fit}
\end{table}

The BESIII measurement gives $|d_\Lambda|<7.0\times10^{-19}\,\ecm$ at
95\% CL~\cite{BESIII:2025EDM}. With the short-distance coefficient this
implies $|d_s|<1.3\times10^{-15}\,\ecm$~\cite{Chen:2026}. The scale variation
in table~\ref{tab:scale} gives bounds in $[0.65,2.3]\times10^{-15}\,\ecm$.
The LCDA input ranges give bounds in $[1.0,1.8]\times10^{-15}\,\ecm$, and
rescattering weakens the central bound by $5\%$ at
$\Lambda_{\text{cut-off}}=1\,\gev$.

Table~\ref{tab:fit} gives the projected standard errors on $|d_s|$ from the
$\Lambda$, $\Sigma^+$, $\Xi^-$ and $\Xi^0$ using the central coefficients.
At BESIII we take the relative precisions of the $\Sigma^+$ and $\Xi$ EDMs
with respect to the $\Lambda$ EDM from ref.~\cite{Guo:2025Polarized}. The
larger $\Xi$ coefficient partly compensates for its lower precision, and
the $\Sigma^+$ constrains $d_s$ mainly through rescattering.
At $\Lambda_{\text{cut-off}}=1\,\gev$, the joint fit lowers the corresponding
$\Lambda$-only upper limits by $27\%$ at BESIII and $26\%$ at STCF. The
projected 95\% upper limits are
\begin{equation}
 |d_s|<1.0^{+0.7}_{-0.5}\times10^{-15}\,\ecm\ \ (\text{BESIII})\qquad
 |d_s|<1.3^{+1.0}_{-0.7}\times10^{-16}\,\ecm\ \ (\text{STCF}).
 \label{eq:joint-bounds}
\end{equation}
The scale variation multiplies the central bounds by factors of $0.49$
to $1.7$.

With the same projected errors, the short-distance fit lowers the
$\Lambda$-only limits by about $21\%$ at either facility. Imposing the SU(6)
ratios of eq.~\eqref{eq:su6-limit} on the final form factors gives
reductions of $18.2\%$ at BESIII and $17.7\%$ at STCF. Relative to the
short-distance joint limits, rescattering shifts the bounds by at most $3\%$
for the cutoffs up to $1\,\gev$ considered here and by $11\%$ at $1.2\,\gev$. We vary the LCDA inputs of
table~\ref{tab:coefficients} at $\Lambda_{\text{cut-off}}=1\,\gev$ and
$\mu=m_{J/\psi}$. The resulting projected 95\% upper limits lie in
$[7.75,13.5]\times10^{-16}\,\ecm$ at BESIII and
$[1.01,1.75]\times10^{-16}\,\ecm$ at STCF. The upper ends of these ranges
weaken the central bounds by up to about one third.

\section{Conclusions}
\label{sec:conclusions}

Using lattice LCDAs, we obtain the leading-twist quark-EDM form factors of
the baryon octet at $m_{J/\psi}$ and reproduce the published $\Lambda$
result~\cite{Chen:2026}. For a strange-quark source the short-distance
ratio $h_{\Sigma s}/h_{\Lambda s}$ lies between $-0.18$ and $-0.04$ over
the lattice input ranges, above the NRQM value $-1/3$. Varying the scale by
a factor of two changes the coefficients by factors of $0.6$--$2$ and the
$\Sigma/\Lambda$ and $\Xi/\Lambda$ ratios by only $10\%$ and $2\%$,
respectively. The first-order shape parameters enhance the $\Lambda$ and
$\Xi$ coefficients by factors of $3.7$ and $2.6$ relative to asymptotic
LCDAs. In the rescattering model at a $1\,\gev$ cutoff, kaon exchange from the
$\Xi$ source accounts for about $80\%$ of the real $\Sigma$ correction. This
makes $|d_\Sigma|$ comparable to $|d_\Lambda|$ and gives the $\Sigma^+$
sensitivity to $d_s$ through rescattering.

The small timelike $\Lambda$ coefficient gives the short-distance BESIII
limit $|d_s|<1.3\times10^{-15}\,\ecm$. This is about three orders of magnitude
weaker than the estimate from the static identification $d_\Lambda=d_s$.
The joint fits with rescattering at a $1\,\gev$ cutoff give projected upper
limits of $1.0\times10^{-15}\,\ecm$ at BESIII and
$1.3\times10^{-16}\,\ecm$ at STCF. The renormalization-scale uncertainty is
about a factor of two. Relative to the short-distance joint limits,
rescattering shifts the bounds by at most $3\%$ for the cutoffs up to $1\,\gev$
considered here and by $11\%$ at $1.2\,\gev$.

\paragraph{Note added.}
While this work was in preparation, ref.~\cite{Mu:2026} appeared. It
computes the quark EDM and CEDM contributions to the $\Xi^-$ and $\Sigma^+$
form factors at leading twist. Its quark-EDM coefficients agree with ours up
to an overall sign, which leaves $|d_B|$ and the $d_s$ limits unchanged.
Rescattering is not considered there.

\acknowledgments

This work is supported in part by the National Natural Science Foundation of China (NSFC) under Grant Nos. 12547104 and
12575096, and by the China Postdoctoral Science Foundation under Grant No. 2025M773361.

\providecommand{\href}[2]{#2}\begingroup\raggedright\endgroup
\end{document}